\documentclass[twocolumn]{aastex7}
\usepackage[T1]{fontenc}

\begin{document}

\title{Strong Polarization Signatures from Magnetically Stabilized Luminous Thin Accretion Disks}

\author[0000-0002-5786-186X]{P. Chris Fragile}
\affiliation{Department of Physics and Astronomy, College of Charleston, 66 George St, Charleston, SC 29424, USA}
\affiliation{Center for Computational Astrophysics, Flatiron Institute, 162 5th Avenue, New York, NY 10010}
\email[show]{fragilep@cofc.edu}

\author{Peter Anninos}
\affiliation{Lawrence Livermore National Laboratory, P.O. Box 808, Livermore, CA 94550, USA}
\email{anninos1@llnl.gov}

\author{Madeline Breiling}
\affiliation{Department of Mathematics, College of Charleston, 66 George St, Charleston, SC 29424, USA}
\email{breilingmg@g.cofc.edu}

\author{Derrick Pickrel}
\affiliation{Lawrence Livermore National Laboratory, P.O. Box 808, Livermore, CA 94550, USA}
\affiliation{Department of Physics, University of California, Merced, 5200 N. Lake Rd, Merced, CA 95343, USA}
\email{pickrel2@llnl.gov}

\author{Gabrielle Ruminski}
\affiliation{St. Paul's School, 325 Pleasant St, Concord, NH 03301, USA}
\email{elleruminski@gmail.com}

\begin{abstract}
We utilize the Monte Carlo radiation transport capabilities of Cosmos++ to explore the polarization and Faraday rotation of radiation emitted from a set of general relativistic radiation magnetohydrodynamic simulations of magnetically stabilized, thin, black hole accretion disks. The guiding question is whether or not the Faraday rotation depolarizes the radiation to such a degree as to be inconsistent with the relatively high polarization measurements coming from the Imaging X-ray Polarimetry Explorer (IXPE). 
After first confirming that our code reproduces expected polarization results for electron clouds, planar scattering atmospheres, and AGN disks, we demonstrate the polarization and Faraday rotation capabilities using a Novikov-Thorne disk threaded with a purely toroidal magnetic field. We then analyze temporal snapshots from three different simulations of luminous, thin accretion disks threaded with different magnetic field configurations. \added{We find that the effects of Faraday rotation are generally modest over the energy range of interest, since the strongest fields present in these simulations ($\gtrsim 10^8$ G) are mostly hidden beneath the photosphere}. We can easily produce polarization degrees $\ge 4$\% for sources seen mostly edge-on. As one example of a relevant target, we compare our results to an IXPE observation of Cyg X-1, finding that we can match the polarization degree quite well, though our polarization angle is rotated approximately $90^\circ$ with respect to the observed data. We speculate on a few possible explanations for this.
\end{abstract}



\section{Introduction} 

Black hole X-ray binaries (BHXRBs) transit through multiple ``states'' during their outbursts. These states are distinguished by their spectral and timing features \citep[see][for reviews]{Homan05, Remillard06}. One of the prominent states, called the ``soft'' or ``thermally dominant'' state, is characterized by a disk-like spectrum (soft, with a prominent thermal bump around 1 keV) and low variability (rms $\lesssim 3$\%). This state is usually observed in a narrow luminosity range around 10-50\% of $L_\mathrm{Edd}$ \citep[e.g.][]{Done07}, where $L_\mathrm{Edd} = 1.2 \times 10^{38} (M/M_\odot)$ erg s$^{-1}$ is the Eddington luminosity for a black hole of mass $M$. 

The source spectrum in the soft state is well described by a multi-temperature blackbody corresponding to the standard thin accretion disk model of \citet{Shakura73}. However, it has been known for decades that the Shakura-Sunyaev solution is thermally \citep{Shakura76}, and possibly viscously \citep{Lightman74}, unstable in its inner region at luminosities above a few percent Eddington, exactly in the range covered by the soft state. This is quite puzzling given the additional association of the soft state with low variability (i.e., the disk appears to be stable).

One proposed explanation for the invariability of disks in the soft state has been to invoke strong magnetic fields as a stabilizing mechanism \citep{Begelman07, Oda09, Sadowski16}. In \citet{Mishra22}, we explored this idea by performing a series of general relativistic radiation MHD (GRRMHD) simulations of luminous, thin accretion disks threaded by different magnetic field topologies. In all cases, the fields were initially weak, with the gas-to-magnetic pressure ratio $\gtrsim 100$. In the end, we found that what we called our dipole and multi-loop configurations were thermally unstable, collapsing on the local thermal timescale. Our vertical case, on the other hand, remained stable for the duration of our test, while our quadrupole case was intermediate, showing signs of both stability and instability. Each of these cases will be described in more detail in Section \ref{sec:sims}. The point is the simulations of \citet{Mishra22} confirmed the hypothesis that strong magnetic fields can support thin disks against collapse.


However, the idea that strong magnetic fields are the solution has recently been challenged. \citet{Barnier24} pointed out that strong magnetic fields in disks may be in conflict with recent high polarization measurements made by the Imaging X-ray Polarimetry Explorer (IXPE) \citep[e.g.,][]{Krawczynski22, Majumder26}. This is because strong magnetic fields in the ambient plasma should lead to large amounts of Faraday rotation of the polarization vector. Other than for certain, very specific field geometries, this Faraday rotation should combine incoherently, leading to a lessening of the overall polarization degree (and\added{ the corresponding polarization transmittance, which is the ratio of the polarization degree with Faraday rotation included to the polarization degree experienced without it}). Since the amount of Faraday rotation increases linearly with the strength of the magnetic field, \citet{Barnier24} \added{suggest} that IXPE measurements effectively place upper limits on the magnetic field strength surrounding black hole accretion disks, and these upper limits may be stringent enough to preclude magnetic fields being the stabilizing mechanism in the soft state.

This conjecture is what we aim to test in the current work. We do this by performing polarized radiative transport, accounting for Faraday rotation, on the very same GRRMHD simulations that were presented as evidence for magnetic stabilization in \citet{Mishra22}. This is the first work that we are aware of that accounts for Faraday rotation in this way using magnetically stabilized luminous thin accretion disks as input. As we will show, the Faraday rotation has only a modest effect on the polarization degree, with the polarization transmittance remaining $\gtrsim 45$\%. Our simulations also had no trouble producing polarization degrees $\ge 4$\% in the IXPE energy range, provided the source is viewed at inclinations $i\gtrsim 60^\circ$.

This paper proceeds as follows. Since this is our first time presenting results of polarized radiative transport using Cosmos++, we take time in Section \ref{sec:methods} to describe how this transport is handled. While we relegate most of our testing and validation to the appendices, we do present a Novikov-Thorne disk example in Section \ref{sec:NT}. With the addition of an ambient magnetic field, this test is particularly useful for demonstrating the effects of Faraday rotation. Then, in Section \ref{sec:results}, we get to the main results of our paper, where we analyze three of the simulations of magnetized, luminous thin accretion disks from \citet{Mishra22}, paying particular attention to how strongly Faraday rotation affects the polarization signatures. In Section \ref{sec:CygX1}, we discuss how our results compare to the well-studied BHXRB Cyg X-1. Finally, in Section \ref{sec:conclusions}, we present some concluding thoughts.

\section{Polarized Radiative Transport with Cosmos++}
\label{sec:methods}

We extend the Monte-Carlo (MC) radiation transport capabilities of Cosmos++ to include the modeling and imaging of {\it polarized} spectral emissions, utilizing the multiple frame approach described in \citet{Roth22, Roth25}.  Essentially, the hydrodynamic and photon geodesic equations are solved in the coordinate frame of the black hole spacetime (in Kerr-Schild coordinates), while the radiation-matter interactions are evaluated in the co-moving fluid frame defined by an orthonormal tetrad attached to the fluid. Additionally, scattering events are evaluated in the electron rest frame by sampling the speed of thermal electrons in the fluid frame from either a Maxwell or Maxwell-Juttner distribution. Given an electron velocity, photons are Lorentz boosted from the fluid frame into the electron frame where the appropriate interaction cross sections for polarized radiation are evaluated via Monte Carlo sampling to derive new scattered energies, propagation directions, and polarization vectors as we now describe. 

In the electron frame, the differential Klein-Nishina cross section for a partially polarized photon beam can be written as a superposition of an unpolarized and completely polarized beam \citep{Matt96, Schnittman13}
\begin{equation}
\frac{d\sigma_\mathrm{KN}}{d\Omega} = (1-\delta)\frac{d\sigma_\mathrm{KN,U}}{d\Omega} + \delta \frac{d\sigma_\mathrm{KN,P}}{d\Omega} ~,
\end{equation}
where $\delta$ is the invariant polarization degree ranging from zero to unity and
\begin{eqnarray}
\frac{d\sigma_\mathrm{KN,U}}{d\Omega} & = & \frac{1}{2} r_0^2 \beta^2 (\beta + \beta^{-1} - \sin^2 \theta) ~, \\
\frac{d\sigma_\mathrm{KN,P}}{d\Omega} & = & \frac{1}{2} r_0^2 \beta^2 (\beta + \beta^{-1} - 2\sin^2 \theta \cos^2 \Phi) ~.
\end{eqnarray}
Here $r_0 = e^2/m_e c^2$ is the classical electron radius, $\theta$ is the scattering angle between the incident and scattered photon directions, $\Phi$ is the azimuthal scattering angle between the polarization vector of the incident photon and the plane of scattering, and $\beta$ is the ratio of outgoing to incident photon energies
\begin{equation}
\beta = \frac{E_1}{E_0} = \frac{1}{1+(E_0/m_e c^2)(1-\cos\theta)} ~.
\end{equation}

The scattering angle $\theta$ is computed using either Khan's rejection method \citep{Kahn54} at photon energies $E<3m_ec^2$ or a direct sampling method \citep{Koblinger75} at higher energies. The azimuthal angle $\Phi$ is sampled either uniformly over $2\pi$ for unpolarized beams, or with the following probability distribution for polarized beams:
\begin{equation}
    p(\Phi)=1-2\delta\left(\frac{\sin^2\theta}{\beta+\beta^{-1}}\right)\cos^2\Phi ~.
\end{equation}
Given the set of scattering angles, the outgoing photon is assigned a direction
\begin{eqnarray}
\mathbf{D} = & \left[\mathbf{D}_0 \cos \theta + \mathbf{P}_0 \sin \theta \cos \Phi \right. \nonumber \\
& \left.+ (\mathbf{D}_0 \times \mathbf{P}_0) \sin \theta \sin \Phi\right]/\vert \mathbf{D}\vert ~,
\end{eqnarray}
given the incident photon direction $\mathbf{D}_0$ and polarization vector $\mathbf{P}_0$. The outgoing photon direction is additionally required to be orthogonal to the outgoing polarization vector $\mathbf{D}\cdot\mathbf{P} = 0$. 

The polarization vector is computed via one of two methods, with the second being used primarily due to its covariant treatment of the polarization 4-vector. The first method evaluates the polarization degree once the scattering angles have been determined \citep{Matt96}:
\begin{equation}
    \Pi_p=2\frac{1-\sin^2\theta \cos^2\Phi}{\beta + \beta^{-1}-2\sin^2\theta \cos^2\Phi} ~.
\end{equation}
If a random number drawn between 0 and 1 is greater than $\Pi_p$ the outgoing polarization vector is chosen randomly in the plane normal to the outgoing photon direction $\mathbf{D}$. Otherwise, it is set by \citep{Angel69}
\begin{equation}
\mathbf{P} = \frac{1}{\vert\mathbf{P}\vert}(\mathbf{P}_0 \times \mathbf{D}) \times \mathbf{D} ~.
\end{equation}

The second method takes advantage of the fact that the polarization 4-vector $P^\alpha$ is defined only to within a multiple of the photon 4-momentum $k^\alpha$. Thus, it can be chosen as \citep{Connors80, Schnittman13}
\begin{equation}
P^\alpha = \left(0, P^i - \frac{P^0 k^i}{k^0}\right) = (0, \cos \psi \mathbf{e}_\parallel + \sin \psi \mathbf{e}_\perp) ~,
\end{equation}
where $\mathbf{e}_\parallel$ is a basis vector in the scattering plane, $\mathbf{e}_\perp$ is a basis vector normal to the scattering plane, and $\psi = (1/2)\tan^{-1} (U/Q)$ is the polarization angle associated with the Stokes parameters $Q = \delta I \cos 2\psi$, $U = \delta I \sin 2\psi$, and intensity $I$. The outgoing (scattered) Stokes parameters are calculated from the Rayleigh phase matrix 
\begin{eqnarray}
Q' & = & I\sin^2\theta - Q(1+\cos^2\theta) ~, \\
U' & = & U(-2\cos\theta) ~, \\
I' & = & I(\beta+\beta^{-1}-\sin^2\theta)-Q\sin^2\theta ~,
\end{eqnarray}
from which the outgoing polarization vector is derived
\begin{eqnarray}
\mathbf{P} & = & \cos \psi^\prime e_\parallel^\prime+\sin \psi^\prime e_\perp^\prime ~, \\
\delta^\prime & = & \frac{\sqrt{Q^{\prime 2}+U^{\prime 2}}}{I^\prime} ~, \\
\psi^\prime & = & \frac{1}{2}\tan^{-1}\frac{U^\prime}{Q^\prime} ~.
\end{eqnarray}
The new (post-scatter) basis frame is similarly defined by a component perpendicular to the scattering plane $e_\perp^\prime=e_\perp$ and an orthogonal component in the scattering plane $e_\parallel^\prime$ subject to the condition $\mathrm{e}_\parallel^\prime \cdot \mathbf{D} = 0$. 

We adopt the standard convention where $\psi = 0^\circ$ corresponds to a polarization vector parallel to the plane of the disk (or equivalently perpendicular to its projected symmetry axis), referring to this as ``horizontal'' polarization. A ``vertical'' polarization angle of $\psi = \pm 90^\circ$ corresponds to a polarization vector parallel to the projected symmetry axis of the disk. Note that this convention only has meaning whenever $i \ne 0^\circ$ and that a ``vertical'' polarization vector will still have a component in the plane of the disk unless $i=90^\circ$.

Photon packets propagate along geodesics between scattering events. During these event-free phases, the polarization vector evolves according to the parallel transport equation $k^\mu\nabla_\mu P^\nu=0$, which is  solved for arbitrary spacetimes in the coordinate frame as
\begin{equation}
\frac{dP^\nu}{dt} = -\Gamma^\nu_{\mu\sigma} P^\sigma k^\mu/k^0 ~.
\end{equation}
This set of first order differential equations are advanced together with the geodesic equations utilizing the same high order Runge-Kutta and Verlet methods described in \citet{Roth22}. We additionally enforce the normalization $P^\mu P_\mu=1$, the orthogonality condition $k^\mu P_\mu=0$, and the rescaling $P^\alpha\to P^\alpha-(P^0/k^0)k^\alpha$ after each solve cycle. 

A typical thin disk calculation requires between 10 and 100 million photon packets to adequately resolve the spectral luminosity attributed to thermal emissions \citep[even more for Compton upscattering;][]{Roth25}. However, due to the characteristically small degrees of polarization observed in our models, we find substantially greater numbers are needed to extract meaningful polarization states from statistical fluctuations. This is a problem especially for polarization angles. Results presented here are derived from samples of more than a billion packets, utilizing roughly $10^4$ CPU-hours of compute time per simulation.

In the appendices of this report, we present results of various test problems as evidence that our polarized radiation transport scheme can reproduce expected results. In Appendix \ref{sec:cloud}, we confirm our code correctly sources photons from ellipsoidal clouds of electrons and tracks their polarization through multiple Thomson scatterings. In Appendix \ref{sec:atmosphere}, we verify our code accurately models multi-temperature, plane parallel scattering atmospheres in both the optically thin and thick limits and at low (Thomson-dominated) and high (Compton-dominated) temperatures. These tests also verify that our code preserves axisymmetry. Appendix \ref{sec:AGN} tests inverse Compton scattering from hot, two-component AGN systems that include recoil effects from reflections between the hot atmosphere and cooler disk.

\subsection{Faraday Rotation}
\label{sec:Faraday}

Faraday rotation occurs when polarized light travels through a magnetized plasma, affecting the propagation speed of the photons and rotating the polarization plane around the photon velocity by an amount (in radians) \citep{Gardner66}
\begin{equation}
\Delta\psi_F=\int \delta\psi_F(s) ds = \frac{e^3\lambda^2}{2\pi m_e^2c^4}\int n_e(s)B_\parallel(s)ds ~,
\label{eqn:Faraday1}
\end{equation}
where $n_e$ is the local electron density and $B_\parallel$ is the magnetic field strength (in Gauss) projected onto the photon's path (could be positive or negative). Implicit in the derivation of (\ref{eqn:Faraday1}) is that electromagnetic frequencies are much higher than the natural oscillation frequencies of atoms, which holds for ideal, perfectly conductive plasmas. 

It is useful for analysis purposes to recast (\ref{eqn:Faraday1}) in terms of the line-of-sight optical depth, $\tau_\mathrm{LoS}$. Provided $B_\parallel$ is constant along the line-of-sight (or understood to be the density-weighted magnetic field projected onto the line of sight), we then have the following simple order-of-magnitude estimate for the Faraday rotation \citep{Barnier24}:
\begin{equation}
\Delta\psi_F= 5.5 \tau_\mathrm{LoS} \left(\frac{B_\parallel}{10^6\,\mathrm{G}}\right)\left(\frac{E}{2.5\,\mathrm{keV}}\right)^{-2} ~\mathrm{degrees}.
\label{eqn:Faraday2}
\end{equation}

Being a simple line integral, $\Delta \psi_F$ is easily accumulated along the path of every photon bundle and tallied at various detector stations in the manner described in \citet{Roth22}. In practice however, $\Delta \psi_F$ can become arbitrarily large if it is tracked throughout the entire domain, particularly for packets buried deep within the photosphere where high densities and strong magnetic fields are common (see Section \ref{sec:NT} for an explanation of how we define the photosphere). We therefore added an option to mask out (ignore and reset) the Faraday integral for bundles passing through the photosphere. This allowed us to evaluate the relative contributions to the net Faraday rotation of the corona separate from the disk body. In practice, we found very similar polarization degrees and angles whether we tracked the Faraday rotation from photon origin or surface of last scattering (SoLS). While the cumulative Faraday rotation within optically thick regions can be quite large, phase wrapping of the polarization angle means the Faraday rotation inside the disk is largely washed out, and all that really matters is the Faraday rotation that is accumulated after the photon's last scattering.  

In addition to simply tracking the total $\Delta \psi_F$ along photon paths, we can also incrementally couple this Faraday effect to the polarization state at each time cycle by applying the differential rotation $\delta\psi_F(s)$ to the polarization vector. Comparing the polarization degrees from calculations run with and without Faraday coupling allows us to compute a transmittance value that is useful for quantifying the effect of magnetic fields on polarized emissions.

\subsection{Spectral Diagnostics and Imaging}

Cosmos++ supports two inline options for extracting diagnostics: 1) an overlaid spherical grid, centered on the black hole, designed to tally and bin outgoing photon bundles onto a five-dimensional spectral grid; and 2) an option to generate two-dimensional projected sky maps using either a backward projection or forward modeling pinhole technique to image the black hole and surrounding disk to a specified field of view. Implicit in both these procedures is the assumption that photons propagate through an optically thin medium to distant observers without additional scattering or Faraday rotation once they have left the simulation grid. 

The first diagnostic option is intended to extract spectra, e.g., X-ray flux, polarization degree, polarization angle, and Faraday angle as functions of energy/frequency over finite intervals of time, polar angle, azimuthal angle, and photon birth region. This procedure is identical to that described in \citet{Roth25}, though extended to include polarization attributes. Because our thin disk models are nearly axisymmetric and we only consider steady-state behavior, the spectral grids in this work are constructed using single bins in azimuth and time. We perform packet-weighted sums over all relevant birth region bins (equally partitioned by radius into 10 distinct regions), and ultimately report our data over 50 energy bins spanning $10^{-2}$ to $10^3$ keV for one or more inclination bins, using a slightly narrower range, $10^{-2}$ - $10^2$ keV, for most of the tests in the Appendices.

The second diagnostic option is intended to create synthetic polarized X-ray images as they would appear to distant telescopes aimed at the black hole center, with user-specified fields of view and telescope placements. The images are constructed in one of two ways: 1) by projecting photon packets backwards along their exit trajectories, until they intersect a planar grid centered on the black hole and oriented perpendicular to the telescope line of sight; or 2) by propagating photons forward through a pinhole aperture from their exit coordinates onto a planar grid centered on the detector. Both methods yield comparable results, but the former generally provides better statistics and is the method by which all image maps are generated in this work. 

Regardless of method, the synthetic image grids are typically composed of $81\times81$ cells covering the field of view from the perspective of telescopes placed at least several times the computational grid radius from the black hole (though we note that statistics, i.e., the number of photon hits, suffers with increasing radius). In this manner we bin the accumulated radiation flux (ergs/s/cm$^2$), and the (intensity weighted) average polarization degree, polarization angle, and Faraday rotation angle. 

We emphasize that these (planar) image grids are constructed independently of the (spherical) spectral grids, and do not by themselves contain spectral information. However, we have designed the detection algorithm to accommodate an arbitrary number of images, filtered by photon energy ranges so that spectral diagnostics can be interpreted through multiple filtered maps.

\section{Novikov-Thorne Disk}
\label{sec:NT}

Before proceeding to analyze our GRRMHD accretion simulations, we consider a simplified test case that is particularly useful for illustrative purposes, that of a Novikov-Thorne (NT) disk \citep{Novikov73}, the relativistic extension of the Shakura-Sunyaev disk \citep{Shakura73}. We choose a near maximally rotating case with black hole spin $a_* = 0.99$, mass $M=10 M_\odot$, disk viscosity coefficient $\alpha_\mathrm{SS}=0.1$, and dimensionless mass accretion rate $\dot{m} = 0.2$ (normalized to the Eddington rate $\dot{M}_\mathrm{Edd} = L_\mathrm{Edd}/c^2 = 1.3 \times 10^{17} M/M_\odot$ g s$^{-1}$). The disk is mapped onto a $168^3$ Cartesian grid covering a distance of $\pm 15~r_g$ in each dimension, where $r_g = GM/c^2$ is the gravitational radius.  We use geometric zoning (decreasing cell dimensions from the outer boundaries to the black hole center by a constant ratio) along all three directions to achieve maximum cell resolutions near the black hole of $0.06~r_g$ along the $x$ and $y$ axes, and $0.006~r_g$ along the vertical $z$ axis (perpendicular to the disk). To facilitate a comparison against a similar test from \citet{Schnittman13}, we do not add a hot corona around the disk. Instead we set the ambient background to a low density, optically thin gas with a temperature of 1 keV. Hence all scattering activity is constrained to the disk body, and Compton effects through the ambient background are negligible. 

This test problem differs from the others presented in the appendices in that here we include general relativistic geodesic transport of photon bundles and allow scattering only within the disk body, not in an atmosphere. Also, like the hot AGN test in Appendix \ref{sec:AGN}, we source photon bundles from inside the disk (100 bundles in each cell per time cycle, or roughly half a billion packets total), but here we do so only along the equatorial plane defined by $z<0.006~r_g$. In this way we initialize thermal photons with zero polarization in the central plane, emit them isotropically (in the rest frame) at a rate proportional to $L_\mathrm{Edd}~ \Delta A$ (where $\Delta A$ is the surface area of the emitting cell, needed to compensate for geometric zoning), then evolve them as they scatter and pick up a net polarization upon exiting the disk. For a sufficiently optically thick disk, this should be roughly equivalent to sourcing polarized, limb-darkened packets from the surface of the disk photosphere \citep[i.e., the technique followed in][]{Schnittman13}.

To test our code's ability to track Faraday rotation, we thread the disk surface and ambient environment with azimuthal (purely toroidal) magnetic fields, with amplitudes in the  $|B| = (5\text{--}50) \times 10^6$ G. The fields are restricted to regions where the optical depth is less than one Rosseland mean free path. The optical depth is found by integrating the scattering opacity from the outer grid boundary downward toward the equatorial plane (perpendicular to the disk), i.e., $\tau_R = \int^0_{z_\mathrm{out}}\rho~\sigma_R~dz$, where $\rho$ is the gas density, $\sigma_R$ the Rosseland opacity, and $z_\mathrm{out}$ is the outer (top or bottom) vertical boundary. Because the accuracy of this procedure is subject to grid resolution ($\tau_R$ can vary substantially across a single cell), we additionally normalize our reported Faraday angles to an effective unit optical depth by dividing them by the mass-weighted average optical depth of all zones containing a non-zero magnetic field (approximately a factor of 2).

\subsection{Polarization}
\label{sec:NT_spectrum}

\begin{figure}
\centering
\includegraphics[width=\linewidth]{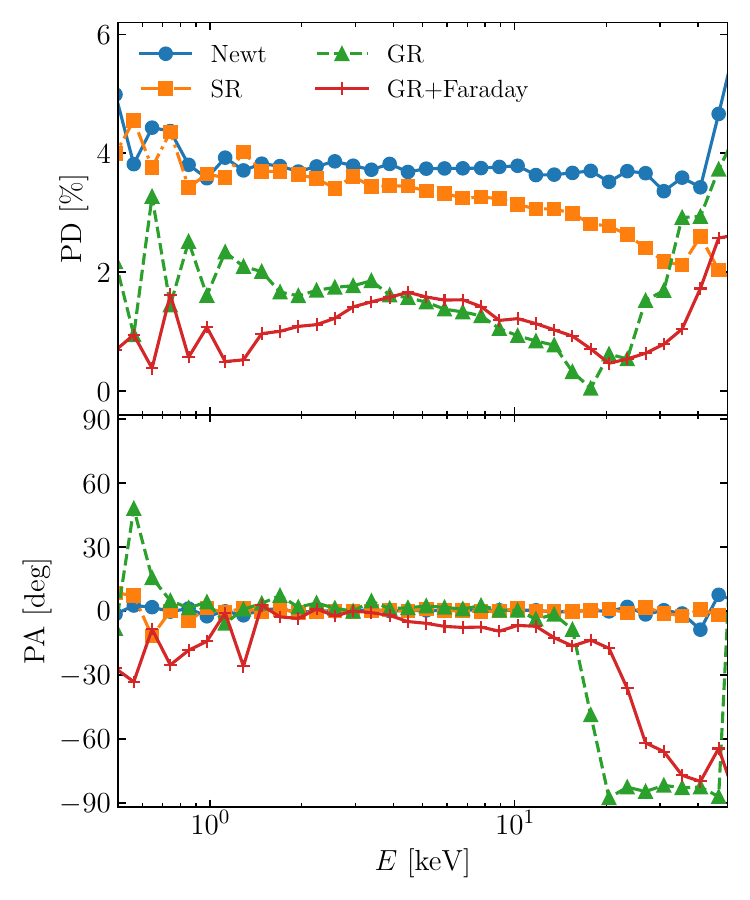}
\caption{Polarization degree and angle for the Novikov-Thorne test case, comparing purely Newtonian, special relativistic, and general relativistic radiative transport both without and with Faraday rotation. All curves are for an observer located at $i=76^\circ$ and magnetic field strength of $10^7$ G.}
\label{fig:NT_spec}
\end{figure}

Figure \ref{fig:NT_spec} shows the polarization degree (PD) and polarization angle (PA) as a function of energy for the NT disk observed at an inclination of $i = 76^\circ$. 
For illustrative purposes, we compare results using Newtonian, special relativistic (SR), and general relativistic (GR) transport, plus a case of GR transport with Faraday rotation accounted for; we will discuss the effects of Faraday rotation in Section \ref{sec:NT_Faraday}. 

The main difference between Newtonian and SR is that the latter properly accounts for Doppler shifts and Doppler beaming. The Doppler effect somewhat reduces the polarization degree (top panel), especially at higher energies. The main difference between SR and GR transport is that GR properly includes the parallel transport of the polarization vector along the photon's geodesic path. Parallel transport causes additional rotation of the polarization vector since it must remain perpendicular to the photon's momentum vector. This leads to a significant reduction in the total polarization degree (top panel of Figure \ref{fig:NT_spec}). In this case ($i=76^\circ$), the rotation associated with parallel transport results in a reduction of the polarization degree from $\approx4$\% to $\approx2$\%, but this effect depends on the inclination of the observer, being stronger for more edge-on viewers. We point out that the overall factor of two greater degree of polarization derived from Newtonian calculations compared to GR ones is in general agreement with earlier estimations \citep[see, for example, Figure 8 from][]{Connors80}.

Another GR effect apparent in Figure \ref{fig:NT_spec} is the change in polarization properties around 20 keV. Here, the polarization degree reaches a local minimum before increasing substantially at higher energies. At the same time, the polarization angle swings from $\approx 0^\circ$ at lower energies to $\approx -90^\circ$ at higher energies. The weakly horizontally polarized photons at low energies come directly from scattering events in the cooler outer parts of the disk, while the more strongly vertically polarized photons that dominate above 20 keV are affected by a number of GR effects, including gravitational lensing, frame dragging by the rotating black hole, and returning radiation\footnote{By returning radiation, we mean photons from one side of the disk traveling over the black hole before striking the opposite side and then reflecting toward the observer.}, all happening in the inner parts of the disk near the black hole. At intermediate energies, the mix of horizontally and vertically polarized photons causes the local minimum in the PD. Where this local minimum occurs depends on the spin of the black hole, but generally occurs near the thermal peak of the spectrum \citep{Schnittman09}.

\begin{figure}
\centering
\includegraphics[width=\linewidth]{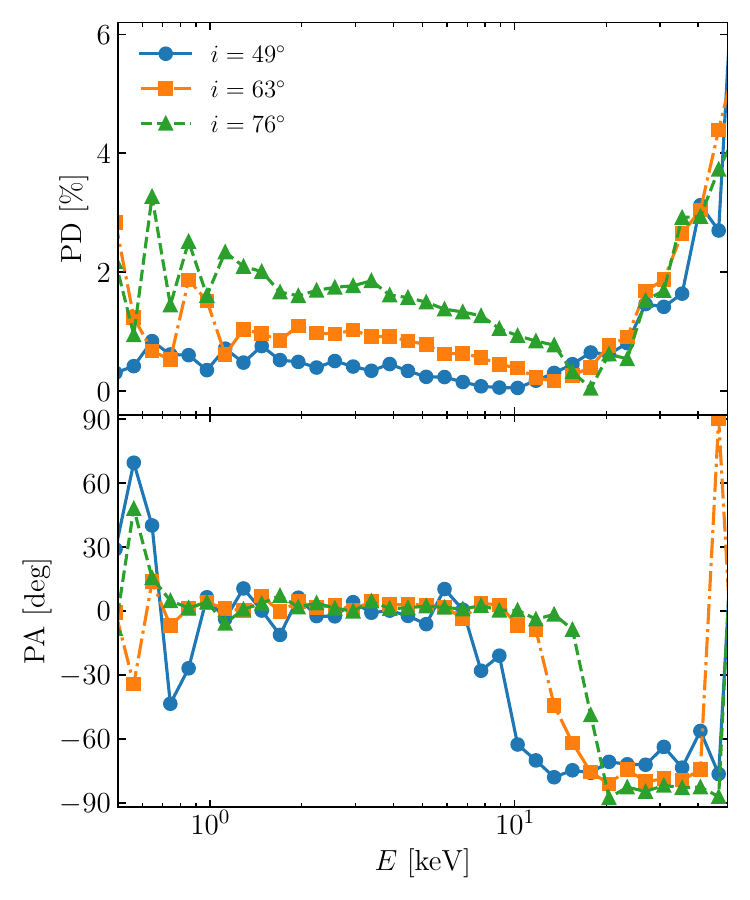}
\caption{Polarization degree and angle for the Novikov-Thorne test case for three different observer inclinations using GR transport, but ignoring the effects of Faraday rotation.}
\label{fig:NT_inclination}
\end{figure}

In Figure \ref{fig:NT_inclination}, we demonstrate that our GR transported polarization properties follow expected dependencies with respect to observer inclination. Below the thermal peak, where photons are coming from the outer parts of the disk, the polarization increases with increasing inclination, as expected for an optically thick, scattering-dominated disk \citep{Chandrasekhar60}. Closer to the black hole, the polarization vectors get rotated mostly in a clockwise direction by various GR effects, leading to a reduction in the polarization degree and negative polarization angles. Then, at even higher energies, the polarization degree picks up again as returning radiation begin to make up more and more of the spectrum \citep{Connors80, Schnittman09}.

\subsection{Synthetic Images}

\begin{figure}
\centering
\includegraphics[width=\linewidth]{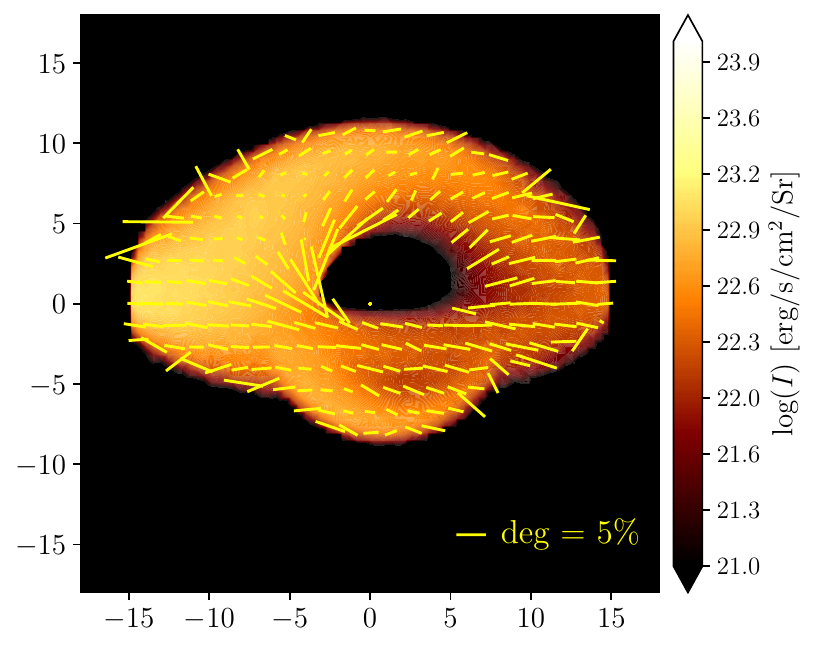}
\includegraphics[width=\linewidth]{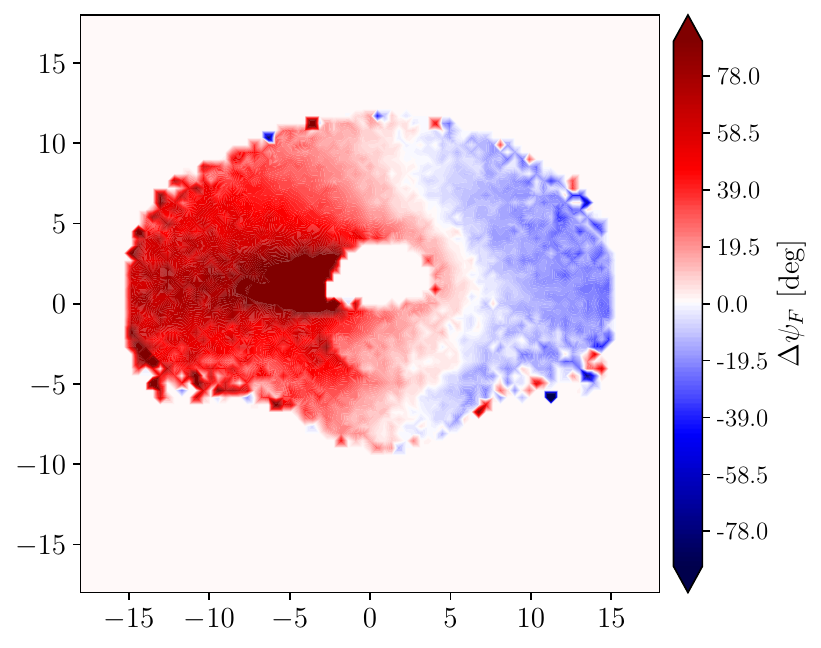}
\caption{Total intensity image with polarization vectors overlaid ({\em top panel}) and a map of Faraday rotation angle ($\Delta \psi_F$) ({\em bottom panel}) for the Novikov-Thorne test case observed at $i=76^\circ$ with a $10^7$ G purely toroidal magnetic field added. The top panel integrates photons over the (2-8 keV) IXPE band, while the bottom panel only includes the 2.5 keV energy bin. Both images represent $30\,r_g \times 30\,r_g$ fields of view centered on the black hole. Photon statistics are incomplete near the outer edges of the disk, leading to spurious results on the fringes of both images.}
\label{fig:NT_image}
\end{figure}

Figure \ref{fig:NT_image} (top panel) shows a total intensity image overlaid with polarization vectors from our NT test disk viewed from close to edge-on ($i = 76^\circ$) with a $30\,r_g \times 30\,r_g$ field of view centered on the black hole. Some of the effects we can see in the image include Doppler beaming, making the left-hand (approaching) side of the disk appear brighter than the right-hand (receding) side, and GR light bending, which warps the back side of the disk into view both above and to a lesser extent below the black hole.

The intensity image is overlaid with polarization vectors, whose lengths represent the PD and whose orientations represent the PA. These vectors are averaged over $3 \times 3$ pixel$^2$ sub-regions in the image to prevent overcrowding. Many interesting effects can be noted. First, the near side of the disk is mostly horizontally polarized, as expected for a planar disk viewed nearly edge on. However, the back side of the disk appears much less polarized because the photons we are seeing from that side departed the disk on much more vertical (low polarization) paths than the ones on the near side, before being bent toward the observer by the gravity of the black hole. 

We also see a difference in the level of polarization between the left and right sides of the disk. In this case, Doppler beaming on the approaching (left) side of the disk bends photons from more vertical trajectories toward the observer's line of sight, again lowering the observed polarization, while the opposite happens on the receding (right) side of the disk, allowing the observer to receive higher polarized photons from that side. Note that the polarization vectors near the outer edge of the disk suffer from poorer photon statistics associated with the truncation of the disk and should mostly be ignored.

\subsection{Faraday Rotation}
\label{sec:NT_Faraday}

\begin{figure}
\centering
\includegraphics[width=\linewidth]{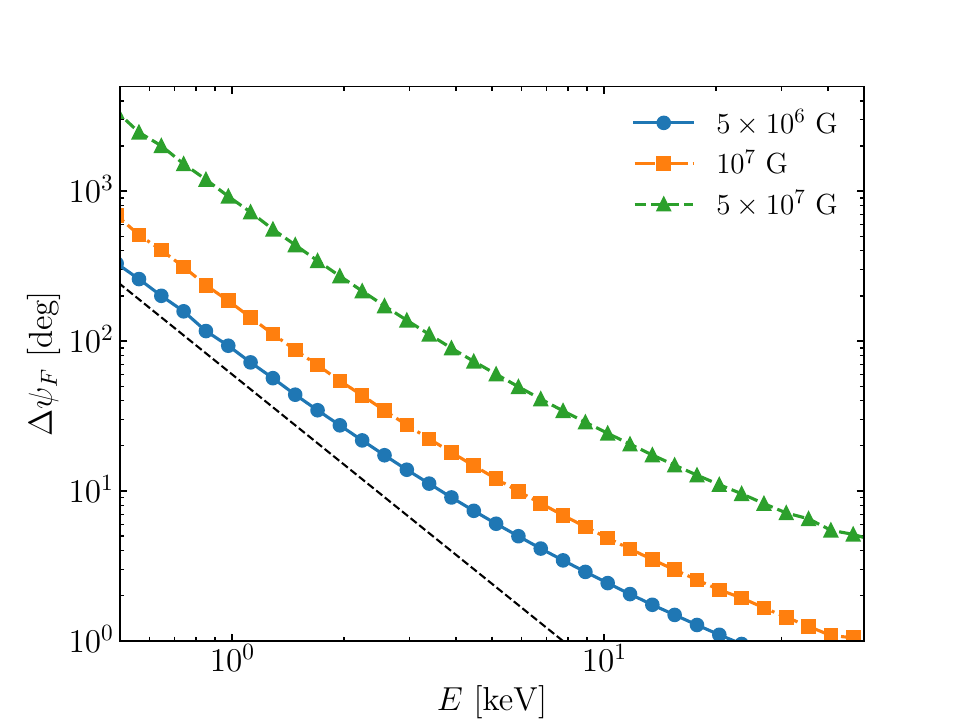}
\caption{Faraday rotation angle as a function of energy for the Novikov-Thorne test with three different magnetic field strengths. The dashed line shows an $E^{-2}$ dependence. This plot is for $i=76^\circ$.}
\label{fig:Faraday}
\end{figure}

We saw in Section \ref{sec:Faraday} that the amount of Faraday rotation depends approximately linearly on the optical depth and strength of the line-of-sight magnetic field component and is inversely proportional to the photon energy squared. These expectations are largely confirmed in Figure \ref{fig:Faraday}, where we plot the integrated Faraday rotation angle $\Delta \psi_F$ as a function of photon energy for three different magnetic field strengths. We see that Faraday rotation is directly proportional to the magnitude of the magnetic field and roughly proportional to $E^{-2}$. There is a slight deviation from the expected dependence at high energies that we attribute to photons that originate close to the black hole\added{. These photons experience larger relativistic corrections, both Doppler shifting and GR effects, that promote greater Faraday rotation. These photons may also} pass through more disk material and magnetic field \added{on their way to the observer} than \added{those} originating further out in the disk, and thus experience a larger product of $\tau_\mathrm{LoS} B_\parallel$.

The bottom panel of Figure \ref{fig:NT_image} provides a map of the Faraday rotation ($\Delta \psi_F$) experienced by photons coming from different parts of the disk for a purely toroidal magnetic field configuration with amplitude $|B| = 10^7$ G, observed nearly edge-on ($i=76^\circ$). At the front and back of the disk, where the field is roughly perpendicular to the line of sight, there is little Faraday rotation ($\Delta \psi_F \approx 0^\circ$). On the left and right sides of the disk, the magnetic field is partially aligned or counter-aligned with the line of sight, producing the largest magnitude Faraday rotation angles. 
The maximum and minimum rotation angles (excluding low-statistics values on the fringes of the disk) are $\Delta\psi_F = +106^\circ$ and $-34^\circ$, respectively, with a total range that is in good agreement with \citet{Barnier24}. There is an overall bias in our Faraday rotation map, favoring more and larger magnitude positive values compared to negative ones\added{. Also, the line of zero Faraday rotation is displaced toward the right side of the disk, and there is a ring of positive Faraday rotation around the black hole. These effects were} not captured in the maps of \citet{Barnier24}\added{, which were made assuming Newtonian physics}. We attribute \added{the bulk of the effect} to Doppler beaming, which boosts more photons into a direction parallel to the magnetic field than anti-parallel to it in this setup \added{where the fluid velocity is parallel to the magnetic field. This provides a bias toward positive Faraday rotation, so even if this disk were viewed face-on ($i=0^\circ$), there would be a net positive Faraday rotation.}

The spatial incoherence of the Faraday rotation seen in Figure \ref{fig:NT_image} is what leads to the additional depolarization of the GR + Faraday spectrum in Figure \ref{fig:NT_spec} relative to the GR spectrum without Faraday rotation. Comparing just these two cases, we first notice that they trace each other rather closely around the emission peak ($\approx 10$ keV)\added{. From Figure \ref{fig:Faraday}, we see that at those energies, we have $20^\circ$ or less Faraday rotation, so the effect on the transmittance is expected to be small. In other words, it makes sense at those energies that the PD would be similar both with and without Faraday rotation. At lower energies, Faraday rotation reduces the polarization degree by about 1\% (in absolute terms) for this particular inclination, field strength, and energy range. This is consistent with the inverse energy dependence of Faraday rotation. The additional} deviation at high energies comes from the complex interaction of Faraday rotation with the many GR effects happening close to the black hole\added{ and the poorer counting statistics at those energies}.  

\begin{figure}
\centering
\includegraphics[width=\linewidth]{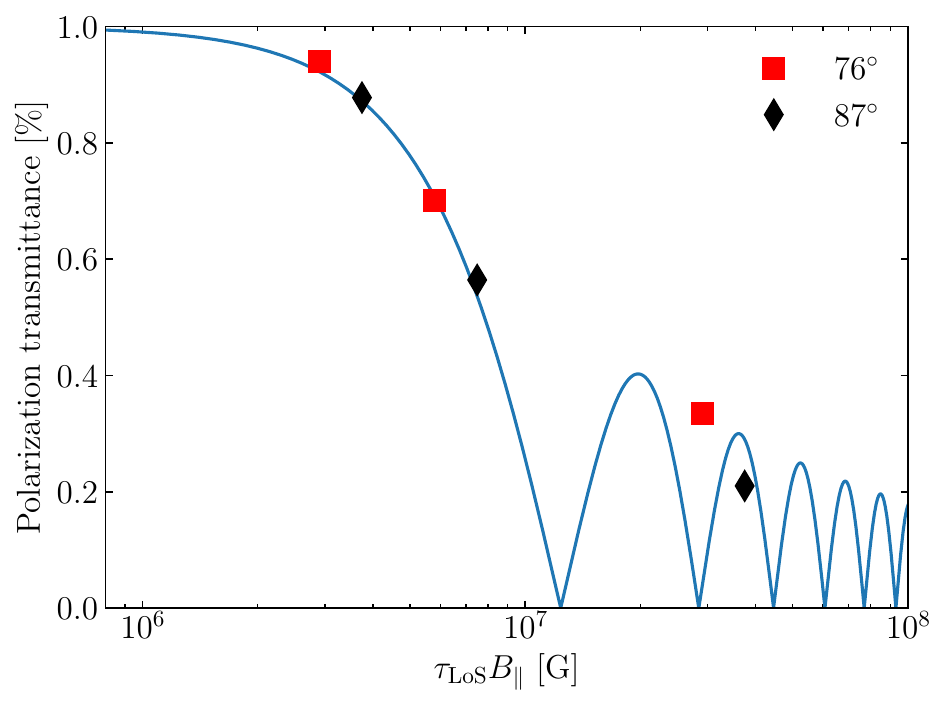}
\caption{A plot of how much the polarization is reduced by the inclusion of Faraday rotation for a purely toroidal magnetic field. The solid line is the prediction from \citet{Barnier24} at 2.5 keV, while the symbols represent results from our Novikov-Thorne test, also centered at 2.5 keV, with field strengths $5\times10^6$ G, $10^7$ G, and $5\times10^7$ G at two different inclinations.}
\label{fig:transmittance}
\end{figure}

The amount of depolarization from Faraday rotation can be better quantified by measuring the polarization transmittance. For weak magnetic fields or lines of sight perpendicular to the field (resulting in $\tau_\mathrm{LoS} B_\parallel \lesssim 10^6$ G), Faraday rotation is negligible and the polarization transmittance is nearly 100\% (see Figure \ref{fig:transmittance}). As $\tau_\mathrm{LoS} B_\parallel$ passes above $10^7$ G, the Faraday rotation for a toroidal field cancels out both the $Q$ and $U$ Stokes parameters over the entire disk and the transmitted signal becomes completely unpolarized. However, because of the phase wrapping of the polarization vector, the polarization transmittance increases again to a secondary (lower) peak at around $\tau_\mathrm{LoS} B_\parallel = 2 \times 10^7$ G, with each successive transmittance envelope getting shorter and narrower \citep[see][]{Barnier24}.

We test this prediction with our magnetized NT disk model by considering field strengths of $B_0 = 5\times10^6$ G, $10^7$ G, and $5\times10^7$ G. By taking the ratio of the polarization degree with Faraday rotation to the polarization degree without it (both at 2.5 keV), we can add data points from our MC transport to the theoretical transmittance plot in Figure \ref{fig:transmittance}. We do this for each of our three field strengths and at two different observer inclinations ($i=76^\circ$ and $87^\circ$) adding six data points to the plot. The agreement with the predicted curve is quite good, confirming that our code matches the expected qualitative and quantitative behaviors of Faraday rotation.

\section{Magnetically Stabilized Luminous Thin Disks} 
\label{sec:results}

Having established that our polarized radiative transport and Faraday rotation methods reproduce relevant test results (Section \ref{sec:NT} and appendices), we now turn to analyzing our simulations of magnetically stabilized disks.

\subsection{The GRRMHD Simulations}
\label{sec:sims}

All of the simulations we discuss were performed on spherical-polar grids using the code Cosmos++ \citep{Anninos05, Fragile14}. The detailed setup of these particular simulations is described in Section 2 of \citet{Mishra22}. All the simulations were initialized from the \citet{Novikov73} generalization of the Shakura-Sunyaev \citep{Shakura73} thin disk with viscosity parameter $\alpha_\mathrm{SS} = 0.02$ and a nominal, target mass accretion rate of $\dot{M}=3L_\mathrm{Edd}/c^2$ (corresponding to $L=0.17L_\mathrm{Edd}$). All of the simulations assumed a non-spinning black hole of mass $M = 6.62 M_\odot$, though we expect the results would apply for any stellar mass black hole in this luminosity range. This exact disk setup was shown in \citet{Fragile18} to be thermally unstable whenever magnetic fields were excluded and an $\alpha$-viscosity was used to provide the necessary angular momentum transport. \citet{Mishra22} then investigated whether magnetic fields could act to stabilize these disks. They considered four magnetic field topologies as we now briefly describe.

The first case (S3Ed) was a zero-net-flux, single-poloidal-loop case. This is the standard dipole field configuration that has been used to initialize many global MHD disk simulations, except in this case the field is much more elongated in the radial direction, extending from near the inner radius of the disk all the way to the outer boundary of the simulation domain. This was to accommodate the very thin nature of the disk and to promote strong radial shear amplification (leading to a growth of the $B^\phi$ component) due to the orbital motion of the disk (the so-called $\Omega$-dynamo), along with amplification from the magneto-rotational instability (MRI). 

The second case (S3Eq) was similar, except instead of a single poloidal loop, it consisted of two poloidal field loops of opposite polarity stacked vertically, one on top of the other, about the midplane of the disk. Again, there is significant field amplification from the orbital motion of the disk. The main difference between this field configuration and the previous one has to do with the location of the initial current sheets. The dipole case (S3Ed) has a single current sheet in the midplane of the disk, whereas the quadrupole case (S3Eq) has two current sheets, each offset, one above and one below the midplane.

The third field configuration (S3Em) consisted of multiple small poloidal loops of alternating polarity distributed in concentric rings moving outward through the disk midplane. Each ring has a width comparable to the local disk height. For such a configuration, the field is unable to amplify significantly due the narrow radial range of each magnetic cell that prevents extensive radial shear. Also, this configuration lacks any sort of underlying guide field that can replenish field lost to reconnection and buoyancy. Ultimately, any amplification of this field is limited to the action of the MRI. 

The final configuration (S3Ev) was a net-flux, vertical-field threading through the disk. Such a field configuration is subject to amplification due to both the shearing at the interface between the disk and background and the MRI inside the disk. 

At the start of all simulations, the magnetic fields were normalized such that $\beta_{\mathrm{mid,}0}=P_{\mathrm{gas,}0}/P_{\mathrm{mag,}0}$ was large and approximately constant with radius. For the dipole, quadrupole, and multi-loop cases, $\beta_{\mathrm{mid,}0} \approx 100$, while for the vertical field, $\beta_{\mathrm{mid,}0} \approx 1000$.

The simulations cover the inner region of the Shakura-Sunyaev disk model from $r=4\,r_g$ to $r=160\,r_g$, 0 to $\pi$ in the $\theta$ direction, and from 0 to $\pi/2$ in the $\phi$ direction, making the simulation domain a wedge shape. As these are very thin disks ($H/R \lesssim 0.03$), a variety of techniques were used to concentrate resolution as much as possible toward the disk, including a logarithmic radial coordinate and a concentrated latitude coordinate.

All the simulations were allowed to run until either the disk collapsed or for a time period of $30,000\,GM/c^3$, the equivalent of over 300 orbits at the ISCO. The most relevant findings of \citet{Mishra22} were: 1) the multi-loop configuration (S3Em) very quickly collapsed (due to thermal instability) and became too thin to be resolved; 2) the dipole configuration (S3Ed) also collapsed, but not completely and still managed to sustain a reasonable accretion rate and luminosity; and 3) the quadrupole (S3Eq) and vertical (S3Ev) configurations stabilized (due to magnetic pressure support). Based on these results, we have decided to focus our analysis on models S3Ed, S3Eq, and S3Ev, while leaving out S3Em. 

Our goal is to determine whether there is significant Faraday rotation in any of these disks that might be detectable by IXPE, the X-ray Polarimeter Satellite \citep[XPoSat;][]{Vatedka25}, or other polarization missions. Specifically, we want to test the prediction of \citet{Barnier24} that this concept of magnetically stabilized luminous thin disks \added{can be} ruled out by the relatively high X-ray polarization measurements of IXPE.

\subsection{The MC radiative transport simulations}
\label{sec:sims_MC}

Because the MC transport code requires Cartesian grids, while the original models were run in spherical-polar coordinates, we first have to map the output (density, temperature, velocity, and magnetic field) from each GRRMHD simulation onto a new Cartesian grid. We use a $192^3$ grid covering the original domain out to $160\,r_g$, geometrically refined to achieve comparable resolution to the GRRMHD calculations in the disk plane and close to the black hole \citep{Roth25}. The finest resolution in each case is $0.06~r_g$ in the $x$ and $y$ directions and $0.0125~r_g$ along the vertical $z$-axis.

The MC treatment of these simulations differs from the relatively simpler treatments of the test problems, which focused on validating particular elements of the code in relative isolation. Here we incorporate a more complete multi-physics treatment by accounting for geodesic transport of the photon packets, polarization by electron scattering, parallel transport of the resulting polarization vector, as well as its Faraday rotation, gravitational redshifts, Compton scattering including photon/electron alignment corrections, relativistic Maxwell-Juttner thermal distributions, relativistic fluid-frame transformations, and free-free emission and absorption opacities. 

Since these disks are much larger than our NT disk model and develop a magnetically threaded coronal atmosphere, we treat every zone in the computational domain as a potential source of radiation. Thus, every cell is allowed to emit 30 initially unpolarized photon packets per time cycle with an isotropic angular distribution specified in the fluid frame, sampling frequencies from the local emissivity function. A typical calculation generates approximately 200 million such photon packets per MC timestep (roughly an ISCO period), and runs for 20 cycles, thus accumulating statistics from over a billion samples.

We note that the original disk simulations extend all the way to the outer boundary (at $r=160 r_g$). In our MC radiative transport, that hard radius cutoff causes unrealistic emission and polarization properties for photon packets originating near that boundary. Furthermore, the simulations were only run for $t \le 30,000 GM/c^3$, so they only had time to reach a quasi-steady state out to a few tens of $r_g$. For these reasons, we choose to only include in our analysis photon packets originating from the first three (of ten total) radial bins, covering $2 \le r/r_g \le 50$. We performed tests using additional radial bins and found our results to be relatively insensitive as long as we did not include the outermost bin. 

\added{We also only analyze a single snapshot from each simulation, taken from their respective end times \citep[see Table 1 of][]{Mishra22}. This obviously ignores any time variability that might have been present in the original simulations. Note, too that IXPE typically requires a very long exposure time ($\sim 10^5$ s) to collect enough photons to make reliable measurements. Thus, our single snapshot is not exactly representative of a typical IXPE observation. However, it would be extraordinarily challenging to match such observations with simulations. First, that timescale is much longer than even the longest duration disk simulations done to date. Furthermore, our MC code requires $\sim 10^4$ CPU hours to process a single snapshot, so there are severe constraints on how many snapshots we could realistically afford to analyze. For now, let us focus on the simpler goals of testing different field geometries and assessing the impacts of Faraday rotation.}

\subsection{Polarization}

Since all of our simulations (S3Ed, S3Eq, and S3Ev) are relatively symmetric about their central axes, we do not expect our results to depend strongly on the observer azimuth, unlike for our previous analysis of tilted, truncated disks \citep{Fragile25}. Plus, the current simulations only covered a $\pi/2$ wedge in the azimuthal direction, so we have forced a certain degree of symmetry in any case. Therefore, as with the other tests presented in this paper, all of the photon packets for each simulation are collected into a single azimuthal bin. Our main focus is then on the energy and inclination dependencies of our results.

\begin{figure}
\centering
\includegraphics[width=\linewidth]{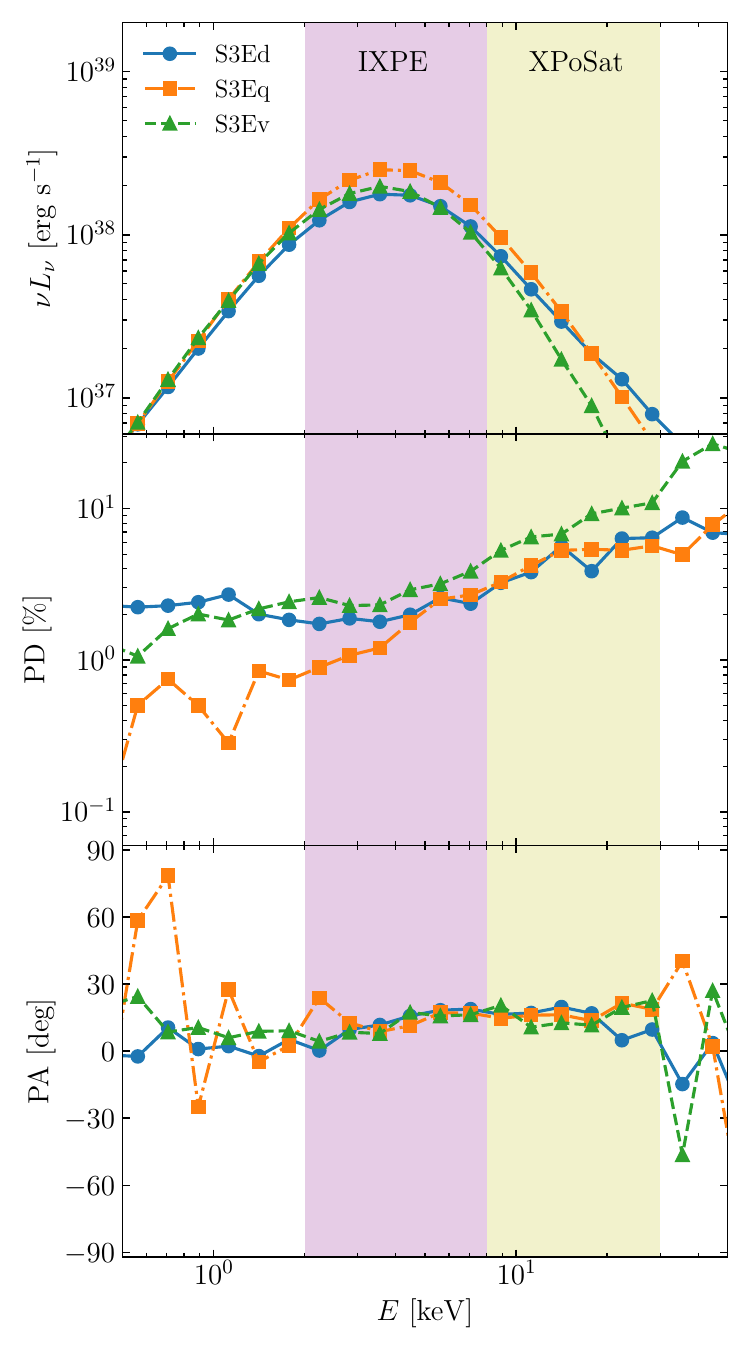}
\caption{Intensity (top panel), polarization degree (middle panel), and polarization angle (bottom panel) plotted as a function of energy for the final dump of all three field configurations. This plot is for an observer inclination of $i = 46^\circ$, and we highlight the sensitivity ranges of IXPE and XPoSat.}
\label{fig:pol_spec_all}
\end{figure}

Figure \ref{fig:pol_spec_all} shows the intensity, polarization degree, and polarization angle for each of the three simulations for an observer at $i = 46^\circ$. In the top panel, the spectra for all three cases look nearly identical. The spectra also agree with the expectations for a system in the soft or thermally dominant state, consistent with the analysis in \citet{Roth25} of a very similar set of simulations. 

Somewhat surprisingly, the three simulations even look fairly similar in their polarization measurements, especially above 4 keV. This would seem to imply that neither the polarization degree nor angle is especially sensitive to the initial magnetic field topology. As we will show in the next subsection, this is consistent with our findings that the Faraday rotation has only a small ($\lesssim 40$\%) effect in this energy range for these simulations. It also suggests that the disk structure is similar enough between the different simulations to be indistinguishable using these spectral measures.  

\begin{figure}
\centering
\includegraphics[width=\linewidth]{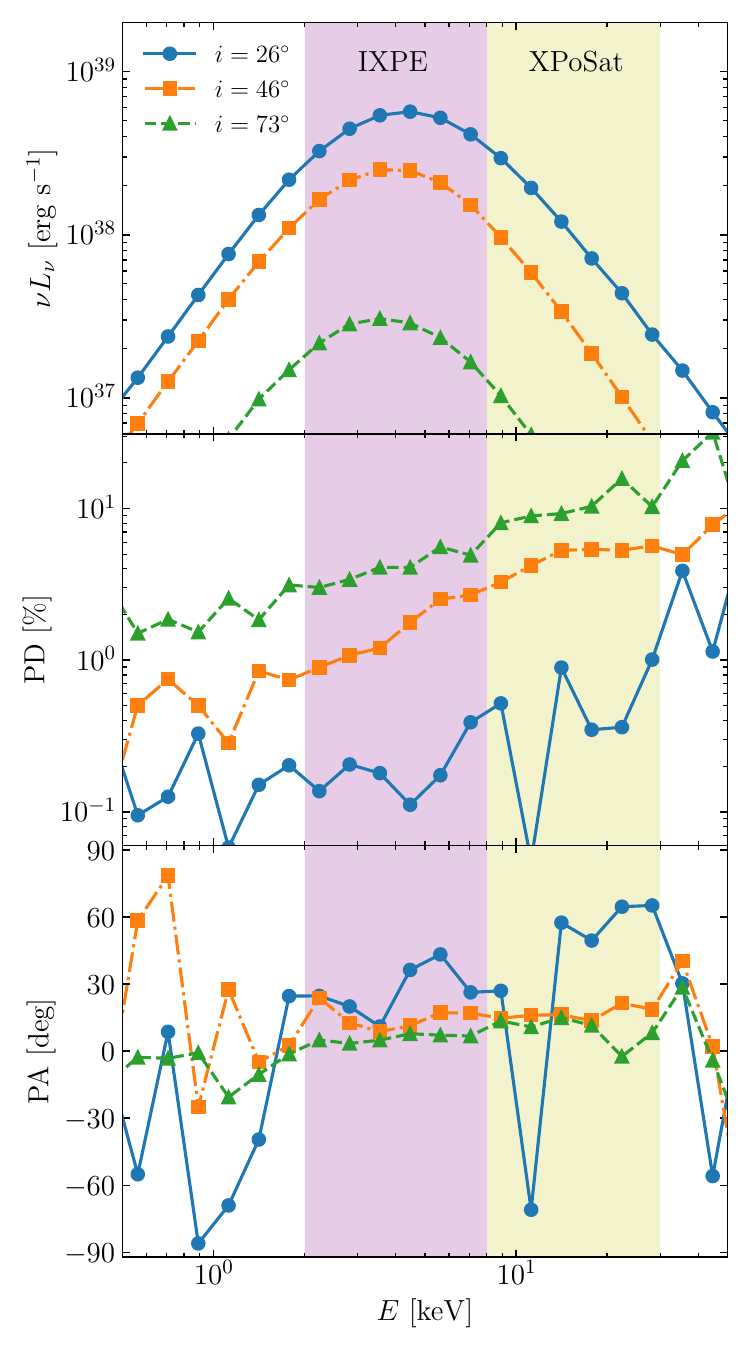}
\caption{Same as Figure \ref{fig:pol_spec_all} except now only for the quadrupole field case (S3Eq) comparing results for different observer inclinations.}
\label{fig:pol_spec_i}
\end{figure}

\begin{figure*}
\centering
\includegraphics[width=\textwidth]{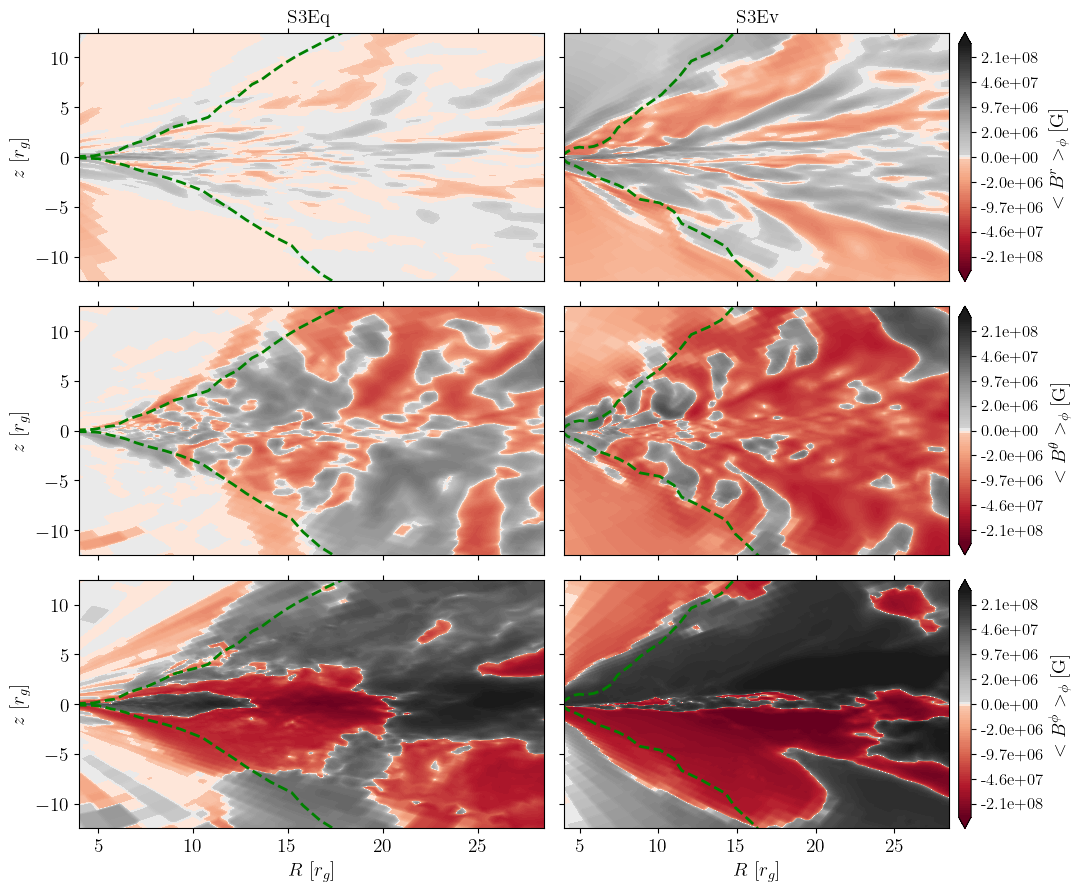}
\caption{Profiles of the azimuthally averaged magnetic field components for the S3Eq (left) and S3Ev (right) cases overlaid with curves representing the $\tau_R = 1$ surface (dashed, green). Data are from the time slices used in our analysis.}
\label{fig:profiles}
\end{figure*}

Focusing now on the quadrupole field case (S3Eq), Figure \ref{fig:pol_spec_i} shows how the intensity, PD, and PA vary with observer inclination. The order-of-magnitude differences in the intensity have to do with the fact that the scattering photosphere is quite thick in these simulations (see Figure \ref{fig:profiles}). An observer at $i\gtrsim 40^\circ$ is looking through this scattering photosphere. Therefore, relatively few of the photon packets originating in the first three radial bins ($r < 50 r_g$) can make it to the observer. However, the polarization degree (middle panel) still behaves as expected, being lowest for the most face-on observer and highest for the most edge-on. The PD can reach as much as 6\% in the IXPE band for an observer at $i=73^\circ$. The PA is close to $0^\circ$ over much of the IXPE and XPoSat bandpasses, consistent with the polarization coming from scattering within the optically thick disk atmosphere. This also suggests GR effects close to the black hole are relatively unimportant. 

\subsection{Faraday Rotation}


In this section, we focus on the quadrupole (S3Eq) and vertical field (S3Ev) cases. Both simulations are ultimately dominated by their azimuthal field component (see Figure \ref{fig:profiles}). For reference, the azimuthal field strength for simulation S3Eq is $B^\phi \approx 3\times10^8$ G inside the disk, about $10^8$ G approximately one density scale-height above it\added{, and $\lesssim 10^7$ G outside the scattering photosphere}. For the S3Ev simulation, the values are $B^\phi \approx 10^8$ G at the disk \added{midplane}, $4\times 10^8$ G just above and below \added{it, and $\lesssim 5\times 10^7$ G outside the photosphere. Depending on the inclination,} these values \added{could be} higher than the upper-limit constraint of $B^\phi \sin i < 8 \times 10^6$ G proposed by \citet{Barnier24} based on the relatively high PD of Cyg X-1. Therefore, we might expect a high degree of Faraday rotation for these simulations, and since the magnetic field ends up being dominated by the azimuthal ($B^\phi$) component, this Faraday rotation should generally lower the PD due to opposite sides of the disk rotating the PA in opposite directions.  

\begin{figure*}
\centering
\includegraphics[width=\linewidth]{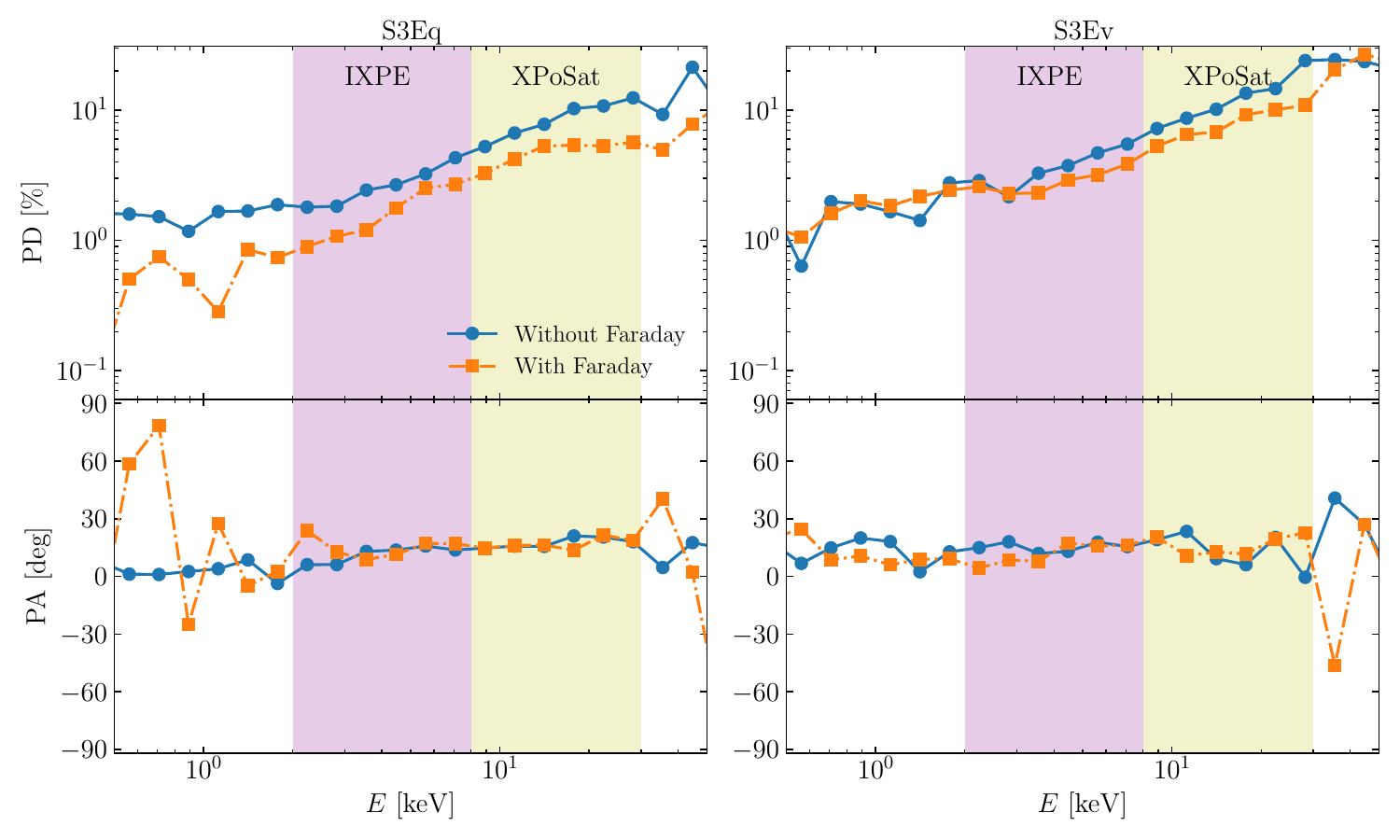}
\caption{Polarization degree (top) and angle (bottom) plotted as a function of energy for the S3Eq (left) and S3Ev (right) cases showing how these quantities compare when Faraday rotation is and is not included. In all cases, the observer is at $i = 46^\circ$. The intensity spectrum is independent of whether or not Faraday rotation is considered, so we do not include a panel for it in this plot.}
\label{fig:Faraday_comparison}
\end{figure*}

The most practical way to measure the impact of Faraday rotation is to directly compare results that do and do not include it, as we have done in Figure \ref{fig:Faraday_comparison}. \added{Apparently,} Faraday rotation has only very modest effects on the S3Eq and S3Ev polarization degrees. As expected, the net result is to generally lower the PD, but the polarized transmittance (ratio of PD with Faraday rotation to PD without it) is only about $0.45$ across the IXPE and XPoSat bandpasses\added{, roughly consistent with the expectations of \citet{Barnier24}, if we only consider the fields strengths outside the surface of last scattering. M}ore surprising is that the transmittance is not strongly energy dependent. Equation (\ref{eqn:Faraday2}) predicts, and we confirmed in Section \ref{sec:NT_Faraday}, that Faraday rotation should go as $E^{-2}$, but such a strong dependence is not apparent in Figure \ref{fig:Faraday_comparison}, at least not above 2 keV. Interestingly, analysis of the total Faraday rotation $\Delta \psi_F$ shows that it does exhibit this energy dependence and confirms that most photons undergo several thousand degrees of rotation in the IXPE range. Yet the cumulative effect is minimal in the polarization spectra of Figure \ref{fig:Faraday_comparison}.

There are at least a couple reasons why the Faraday rotation is less prominent in these simulations than one might expect based upon our NT test results. First, as mentioned previously, almost all of the Faraday rotation in this case is happening {\em inside} the optically thick disk, whereas for the NT test, the field was specifically restricted to lie {\em outside} this region (only regions with $\tau_R < 1$ had magnetic field). The second reason is that the magnetic field strength is much less uniform in the GRRMHD simulations than it was in the NT test. While the azimuthal field component dominates, the radial and vertical field components can be locally strong as well, making the effects of Faraday rotation much harder to predict.

\section{Comparison to Cygnus X-1}
\label{sec:CygX1}

Cygnus X-1 (Cyg X-1) is a high-mass X-ray binary source composed of a stellar mass black hole of $21.2\pm2.2\,M_\odot$ in a 5.6-day orbit with a $40.6^{+7.7}_{-7.1}\,M_\odot$ blue supergiant companion \citep{Krawczynski22}. Winds from the companion star feed an accretion disk around the black hole, which is the source of strong X-ray emission. A pair of radio jets, presumed to lie perpendicular to the plane of the accretion disk, are also often associated with Cyg X-1.

The presence of radio jets has led to the assumption that the accretion disk and black hole must be threaded by substantial magnetic fields based on the current understanding of the physics of jet launching. Thus, it is plausible that the polarization signal from Cyg X-1 should experience appreciable Faraday rotation \citep{Barnier24}, which may be in conflict with the relatively high polarization degree and largely energy-independent polarization angle measured for this source \citep{Krawczynski22, Steiner24, Kravtsov25}. 

\begin{figure}
\centering
\includegraphics[width=\linewidth]{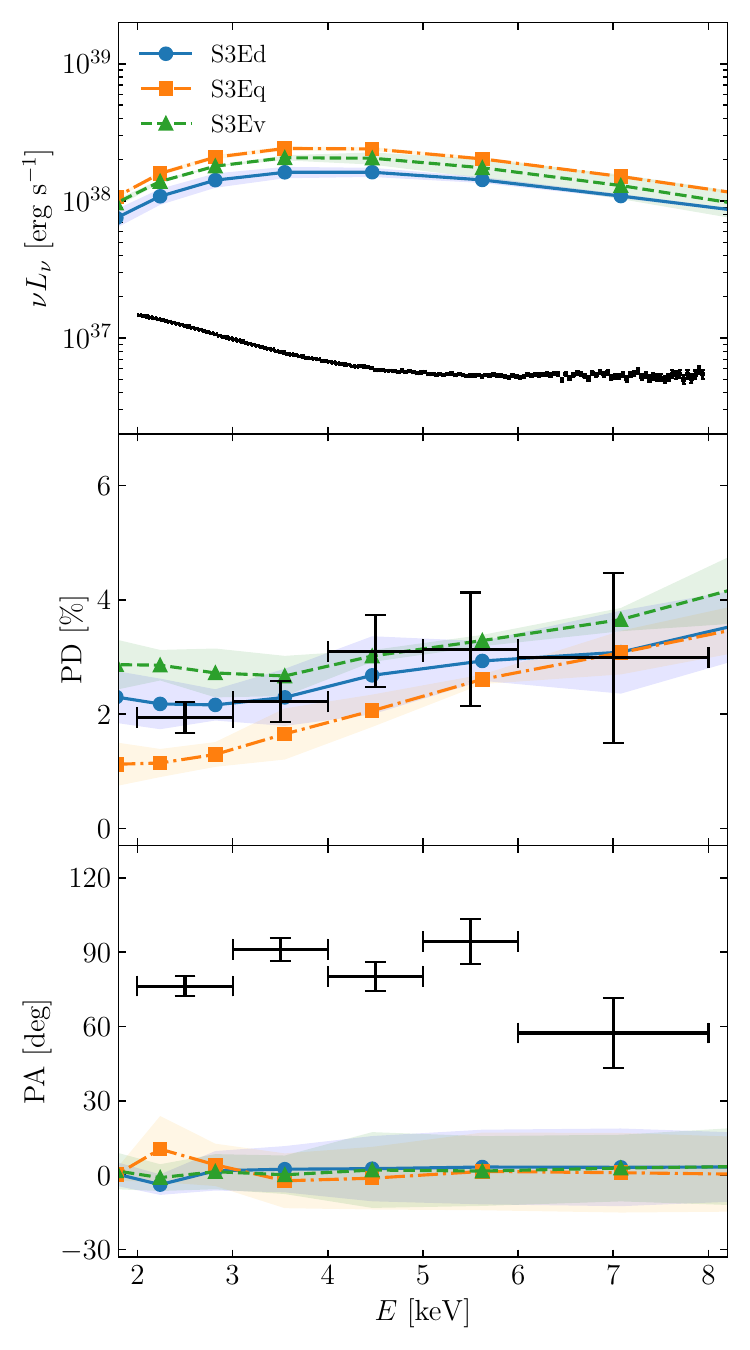}
\caption{IXPE observation of Cyg X-1 (Obs. ID 02008601; black data) overlaid with the \added{intensity (top panel)}, polarization degree (middle panel), and polarization angle (bottom panel) from our three MHD models with $i=46^\circ$. The spectrum for Cyg X-1 assumes a distance of 2.2 kpc, and the PA has been adjusted such that $90^\circ$ would be aligned with that system's radio jet. The shaded regions give some idea of the statistical uncertainties in our MC transport.}
\label{fig:CygX1}
\end{figure}

Since a purely vertical magnetic field only rotates the polarization angle without decreasing the polarization degree, such a field configuration could, in principle, preserve the relatively high PD of Cyg X-1. However, such a field would still introduce an {\em energy-dependent} shift in the PA, which is not seen in the Cyg X-1 data (bottom panel of Figure \ref{fig:CygX1}). This led \citet{Barnier24} to place an upper limit of $B^z < 2\times10^6$ G on the vertical field component. 

An important constraint from the Cyg X-1 data is that its polarization angle aligns roughly with the position angle of its jet throughout the respective IXPE campaigns \citep{Krawczynski22, Kravtsov25, Majumder26}. This suggests that the PA may be oriented along the symmetry axis of the disk. For our simulations, this would correspond to $\mathrm{PA} \approx \pm90^\circ$.

As an illustrative example, we attempt to fit the 20 June 2023 IXPE observation of Cyg X-1 (ID 02008601)\added{. For the polarization, our only free parameters are the initial magnetic field topology of the simulation and the inclination angle of the observer. For} all three of our magnetic field configurations, we find, similar to \citet{Krawczynski22}, that we cannot fit the PD with the $27.5\pm0.8^\circ$ inclination inferred from optical observations of the binary \citep{Miller-Jones21}. Instead, we require an inclination of $i\gtrsim46^\circ$. In other words, the X-ray emitting region must be seen more edge-on than the binary orbit, indicating the disk could be warped.

Our best fit results for the PD are shown in Figure \ref{fig:CygX1} (\added{middle} panel). Using $i=46^\circ$, we found that all three simulations yielded PD of 1-4\%, with a slight, positive energy dependence. Since our disks should look nearly the same from their tops and bottoms, we paired our inclination bins symmetrically about the disk midplanes to give us some idea of our statistical uncertainties (shaded regions in Figure \ref{fig:CygX1}). The S3Ed data fits the Cyg X-1 data the best across the IXPE band, but all three simulations provide reasonable agreement, given our crude inclination binning.

Although Figure \ref{fig:CygX1} confirms that our models can, in principle, reproduce the high PD seen from Cyg X-1, \added{our simulations do not match the spectral shape or PA of that source. Our disk temperatures appear to be too hot, with spectra peaking at around 4 keV, instead of 1 keV or less. We also} get the wrong orientation for the PA. Our polarization angles are much closer to $0^\circ$ (meaning essentially parallel to the disk plane) than to $\pm 90^\circ$ as required by the Cyg X-1 data.  

One way to rotate the polarization vector perpendicular to the disk plane would be through GR effects close to the black hole, especially returning radiation (see Section \ref{sec:NT_spectrum}). However, returning radiation requires a rapidly spinning black hole in order to dominate the PA. Since our simulations are for non-spinning black holes, we find this effect to be weak. Furthermore, there are arguments that returning radiation and a high spin may be incompatible with Cyg X-1, as they would normally introduce a strong energy dependence to the PA, which is not seen \citep{Niedzwiecki26}. An optically thin ($\tau \le 2$) slab corona could also rotate the PA by $\pm90^\circ$ \citep[see Appendix \ref{sec:atmosphere} and][]{Schnittman10}, but such a corona is not formed in our strongly magnetized thin disk simulations. A more radical possibility is that the jet in Cyg X-1 is not, in fact, perpendicular to its disk.



As a reminder, a big part of our motivation for the present work was to test the prediction by \citet{Barnier24} that strong magnetic fields in Cyg X-1 (whether in its disk or its jet) should lead to significant Faraday rotation. Given this, we reemphasize that the effects of Faraday rotation were rather modest for our simulations. Its inclusion resulted in a relatively energy independent drop in the PD of $\lesssim 55$\%, with no discernible impact to the PA. Since all of our models are able to produce polarization degrees of $\ge1$\% provided $i\gtrsim 46^\circ$, it would seem the relatively high PD of Cyg X-1 does not definitively rule out strongly magnetized accretion disks. The work of \citet{Moscibrodzka24} argues further that the presence of prominent jets also does not preclude high polarization.

\section{Conclusions} 
\label{sec:conclusions}

This paper introduces a new capability of our Cosmos++ Monte Carlo transport code \citep{Roth22,Roth25} for tracking polarization, including Faraday rotation effects. With the launch of IXPE, XPoSat, and hopefully continuing with proposed future missions such as the enhanced X-ray Timing and Polarimetry mission \citep[eXTP;][]{Zhang19}, polarization studies of black hole accretion disks are now possible in X-rays. Among other potential discoveries, these missions can probe the geometries of accretion disks \citep{Schnittman09, Cheng16, Farinelli23} and coronae \citep{Schnittman10, Marinucci22, Gianolli23}, as well as constrain the black hole spin parameter \citep{Marra24}. This has motivated new work to try to connect GRRMHD simulations to real systems through comparisons of polarization properties \citep[see][for a review]{Fragile26}.

In Section \ref{sec:NT} and appendices, we demonstrated that this new capability can successfully track polarized radiation through static clouds, planar atmospheres, accretion disks, and in curved spacetimes of highly rotating black holes. The addition (and testing in Section \ref{sec:NT_Faraday}) of Faraday rotation is potentially important as many black hole accretion disks are expected to harbor strong magnetic fields that could dramatically change their polarization signatures \citep{Barnier24}. 

Despite the expectation that Faraday rotation should be highly sensitive to the magnetic field strength and orientation, we found very little difference in its effects among our simulations, despite them having three very different initial field topologies. This could either be because the dominant field component ended up being the azimuthal one in all three cases, or because the field was mostly concentrated inside the optically thick disk, with much weaker values outside.

Not only were the effects of Faraday rotation not very sensitive to our field topology, but the overall effect was rather weak. Near the peaks of the spectra (around 4 keV), Faraday rotation reduced the PD by no more than 55\% and had hardly any effect on the PA. Even with Faraday rotation included, our simulations had relatively little trouble producing polarization degrees of $\gtrsim 4$\% for high inclinations. We conclude therefore that the inclusion of Faraday rotation does not kill the idea of strong magnetic fields stabilizing luminous thin accretion disks. Strong magnetic fields could still potentially be problematic in the hard state since more of the plasma will be optically thin and it may no longer be possible to suppress the impact of Faraday rotation.

To give a specific comparison, we fit our GRRMHD simulation results to an IXPE observation of the BHXRB Cyg X-1 in its soft state. Our simulations rather easily reproduced the observed high PD, provided the inclination angle of the X-ray emitting region was allowed to be $i\gtrsim 46^\circ$. The only issue was that our simulations were off by almost $90^\circ$ on the PA, with ours parallel to the disk plane and the Cyg X-1 PA aligned with the jet axis. One plausible explanation for this discrepancy is that returning radiation is more prominent in Cyg X-1 than in our simulations, perhaps because of a higher black hole spin \citep[see][]{Steiner24, Niedzwiecki26}. Another possibility is that a modest slab corona, which was absent from our simulations, may be required to fit the polarization of Cyg X-1. 

\begin{acknowledgments}
\added{We would like to thank Noemi Barnier and Chris Done for their early reading and feedback on this work, the anonymous referee for their helpful comments and suggestions, and Anastasiia Bocharova, Alexandra Veledina, and Adam Ingram for providing help with the Cyg X-1 IXPE data. }This work was performed under the auspices of the U.S. Department of Energy by Lawrence Livermore National Laboratory under contract DE-AC52-07NA27344. PCF gratefully acknowledges support from NASA under award No 80NSSC24K0900. D. P. was supported by the U.S. Department of Energy, National Nuclear Security Administration, Minority Serving Institution Partnership Program, under Award DE-NA0003984. The Flatiron Institute is a division of the Simons Foundation.
\end{acknowledgments}

%


\software{Cosmos++ \citep{Anninos05, Fragile14, Anninos20, Roth22}}

\appendix

\section{Ellipsoidal Cloud}
\label{sec:cloud}

\begin{figure}
\centering
\includegraphics[width=0.5\linewidth]{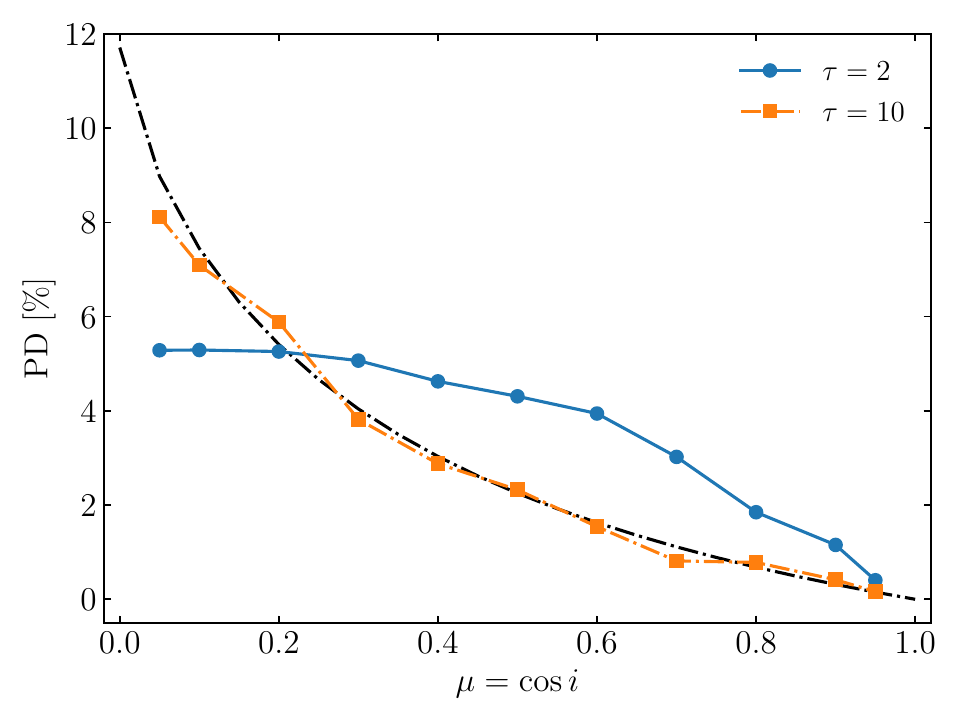}
\caption{Polarization degree PD as a function of inclination $i$ for a uniformly emitting spheroid with $c/a =1/3$ and $\tau = 2$ or $\tau = 10$. The latter very nearly reproduces the \citet{Chandrasekhar60} result for a scattering-dominated disk atmosphere (black dot-dashed line). Compare to Fig. 2 and 4 of \citet{Angel69}.}
\label{fig:Angel}
\end{figure}

This first calculation, attributed to \citet{Angel69}, tests the polarization at various inclination angles $i$ from an oblate spheroidal source with semi-major axis $a$ and semi-minor axis $c$, in which X-rays are uniformly generated and scattered without absorption. Figure \ref{fig:Angel} shows the resulting average polarization degree at eleven inclinations ($\mu = \cos i$) for clouds of optical depth $\tau=2$ and 10, tracking several billion photon packets. When looking at these cloud face-on ($\mu = 1$), the polarization degree is negligible, consistent with its axisymmetric geometry. The polarization degree grows to $\gtrsim 5$\% when viewed edge-on ($\mu=0$). We do not show the result, but we get identical values for $Q/I$ and PD for each angular bin, confirming that our code preserves axisymmetry (i.e., there are no spurious introductions of $U$). In the optically thick ($\tau \gg 1$) limit, the polarization approaches the expected solution for a scattering-dominated medium \citep{Chandrasekhar60}. These results compare nicely with Figures 2 and 4 of \citet{Angel69}.

The fact that Thomson scattering through ellipsoidal clouds of electrons can yield such high degrees of polarization may be relevant for understanding polarized emissions from IXPE sources. As we shall see in the next appendix, the corresponding polarization angle PA may be an important diagnostic for confirming whether or not the corona, which can be effectively modeled as an ellipsoidal cloud, is the source of polarization.

\section{Planar Scattering Atmospheres}
\label{sec:atmosphere}

Our next set of tests looks at the polarization emerging from planar scattering atmospheres, representing black hole accretion disks sandwiched by slab coronae. The disk is the source of photons, which develop a net polarization as they propagate and scatter through the corona on their way to a distant (off-grid) observer. The initial radiation coming from the disk is unpolarized and emitted isotropically as a multi-temperature, multi-frequency blackbody source with a radially dependent temperature profile appropriate for geometrically thin disks \citep{Shakura73}. For the corona we consider two different temperatures: $kT_e = 1$ keV, for which only Thomson scattering is relevant, and $kT_e = 100$ keV, where Compton scattering is important. These tests do not include any of the curvature effects associated with general relativity, and do not actually model the disk. Photons are instead sourced along the equatorial plane so that only scattering through the atmosphere (on one side of the disk) is considered. For these tests we track roughly 50 million photon packets.

\begin{figure}
\centering
\includegraphics[width=\linewidth]{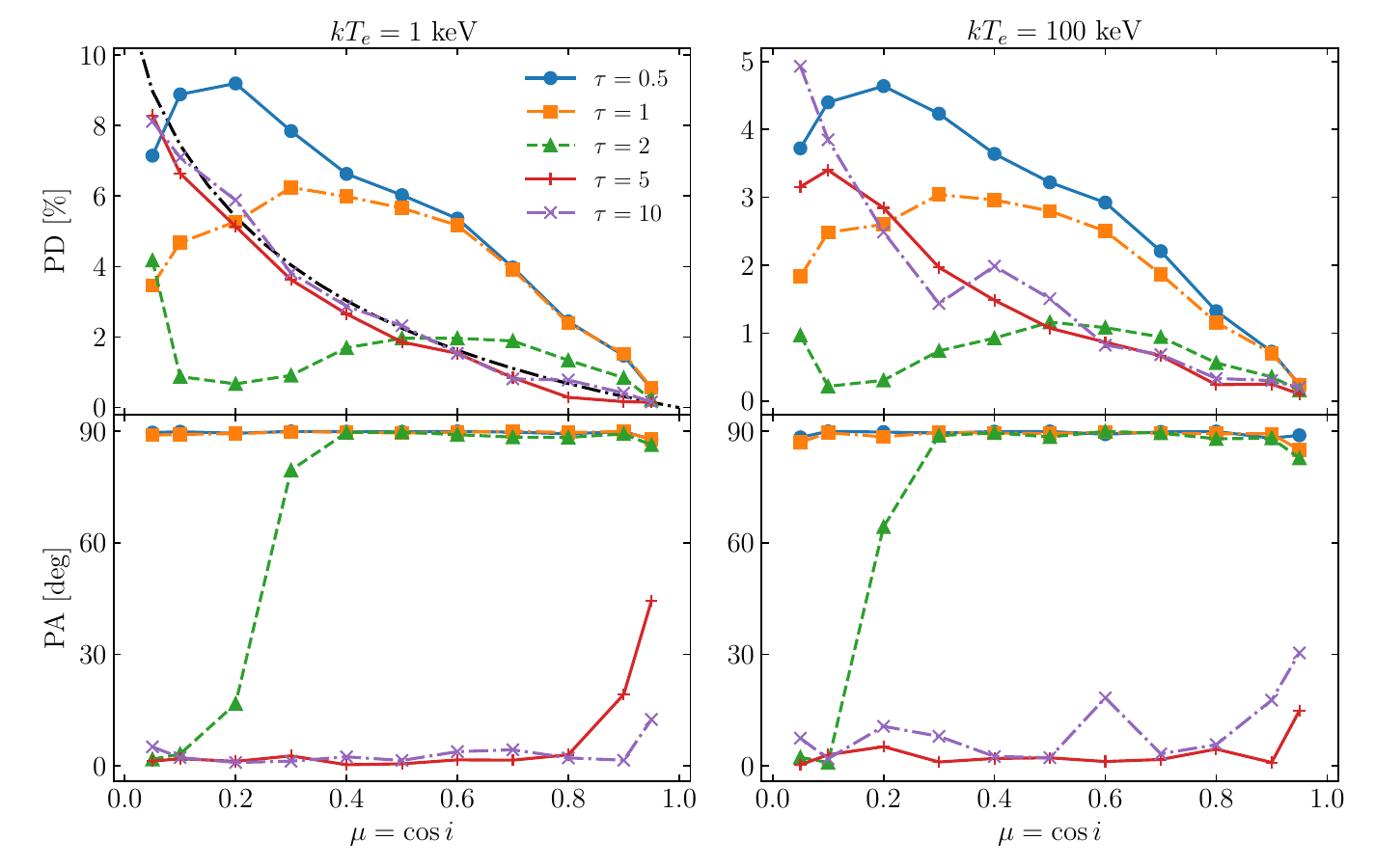}
\caption{Polarization degree (upper panels) and angle (lower panels) as a function of inclination $i$ for different values of $\tau$ for a slab corona with $k T_e = 1$ keV (left column) and $k T_e = 100$ keV (right column). The two highest optical depth cases in the upper-left panel match the \citet{Chandrasekhar60} solution (black dot-dashed line). These results compare nicely with Figures 8 and 9 of \citet{Tamborra18}, across all optical depths.}
\label{fig:Tamborra}
\end{figure}

Figure \ref{fig:Tamborra} shows the polarization degree and angle as a function of inclination ($\mu = \cos i$) for both models, at optical depths ranging from $\tau = 0.5$ (optically thin) to $\tau = 10$ (optically thick). The polarization degree and angle are averaged over all 50 frequency bins at each inclination, reproducing (and, in fact, comparing nicely with) two similar tests from \citet{Tamborra18} (see their Figures 8 and 9). The low temperature case was also presented in earlier work by \citet{Dovciak08}. 

As with the ellipsoidal cloud in the previous appendix, these atmospheres exhibit negligible polarization when viewed face-on ($\mu = 1$), again consistent with their axisymmetric geometry. At intermediate inclinations, the polarization degree is dependent upon optical depth. When the optical depth is low ($\tau \le 1$), the polarization degree is relatively high and the polarization angle is vertical ($\mathrm{PA} = 90^\circ$). As the atmosphere becomes optically thick, we see a transition from vertical to horizontal polarization ($\mathrm{PA} = 0^\circ$), starting from high inclinations ($\mu \lesssim 0.3$) at $\tau=2$ to almost entirely horizontal at all inclinations at $\tau=10$. The reason for this transition can be easily understood. In the limit of low density (low optical depth), only photons traveling in a direction roughly parallel to the plane of the disk pass through enough of the atmosphere to be likely to scatter. In this case, the electric vector after scattering will be predominantly perpendicular to the plane of the disk (hence, a PA of $\pm90^\circ$). In the opposite limit of high density (high optical depth), photons traveling parallel to the disk are unlikely to escape. Instead, only those photons that are able to diffuse up to the surface can escape, and their most likely direction of travel prior to scattering will be normal to the surface. After scattering, the electric vector for these photons will predominantly be in the plane of the disk with $\mathrm{PA} \approx 0^\circ$. 

At our highest optical depths ($\tau \ge 5$), the results for the cooler atmosphere ($kT_e = 1$ keV) begin to resemble the \citet{Chandrasekhar60} solution for an optically thick, scattering-dominated disk, with horizontal polarization increasing from 0\% (at $\mu=1$) to $\sim 12$\% (at $\mu=0$). The high-temperature atmosphere ($kT_e = 100$ keV) demonstrates the effect of Klein-Nishina scattering, which reduces the overall degree of polarization by about a factor of 2 at all energies compared to pure Thomson scattering.

\section{Hot AGN}
\label{sec:AGN}

The final two problems, taken from \citet{Poutanen96} \citep[also presented in][]{Schnittman13}, are intended to test inverse Compton effects from a hot corona, and recoil contributions from reflections off the cooler disk. The two problems differ in corona temperature and optical depth: the first represents a relatively cold atmosphere ($\Theta=kT_e/m_e c^2 = 0.11$) at high optical depth ($\tau = 0.5$); the second is a hotter ($\Theta=0.69$) but less opaque ($\tau=0.05$) environment. These tests differ from Appendix \ref{sec:atmosphere} in that here we model a two-component system by explicitly including the cold disk together with the corona in order to account for Compton recoil effects. Both the disk and corona are treated as uniform slabs with vertical scale heights set by their respective optical depths (the optical depth of the disk is arbitrarily assigned to be about 100 times the depth of the corona). In both cases the disk is assumed to emit unpolarized thermal radiation at 10 eV uniformly throughout the disk, while sampling a relativistic Maxwellian distribution at each scattering event.

\begin{figure}
\centering
\includegraphics[width=\linewidth]{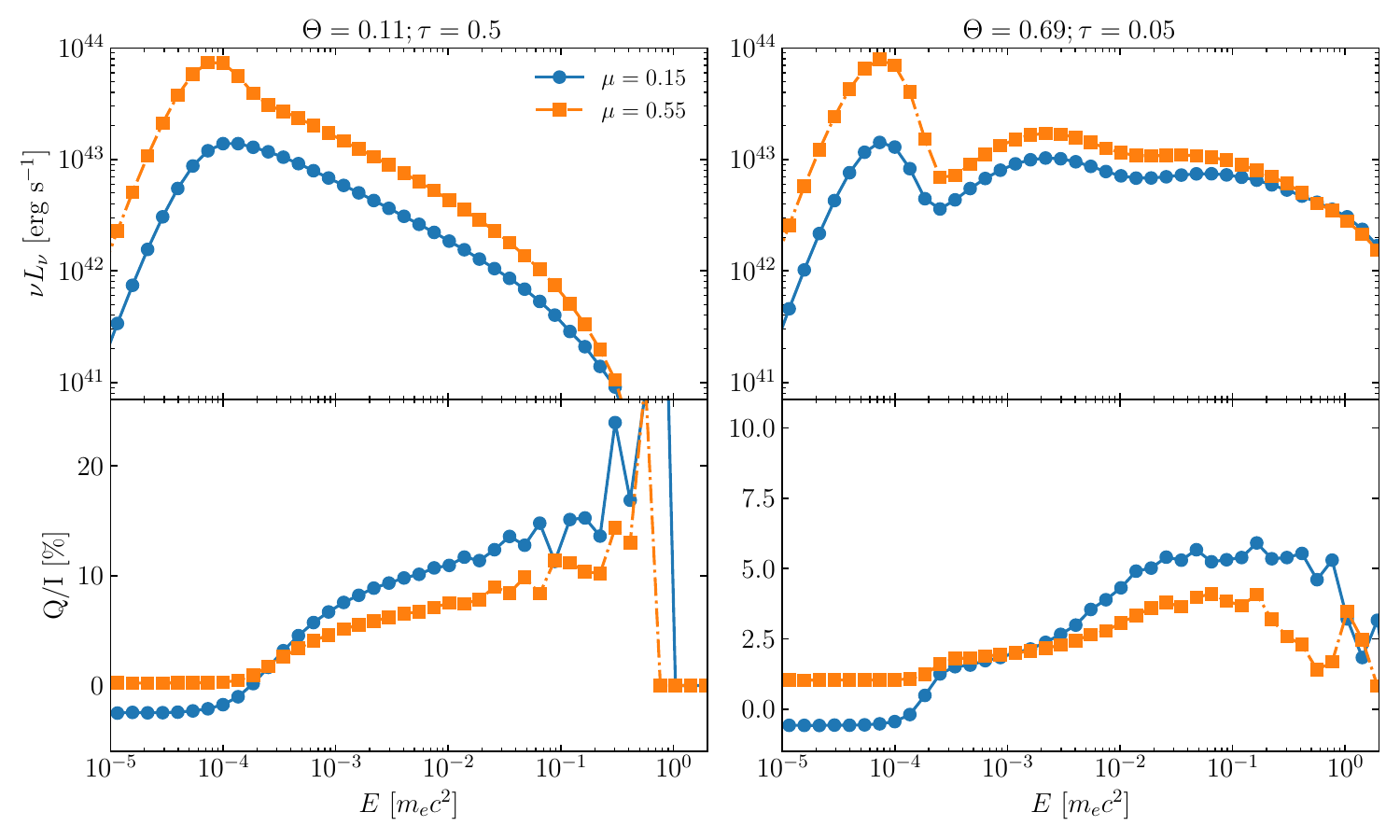}
\caption{Flux (upper panels) and ratio $Q/I$ (lower panels) as a function of photon energy for a relatively cold, optically thick corona ($\Theta = 0.11$, $\tau = 0.5$; left panels) and a hot, optically thin corona ($\Theta = 0.69$, $\tau = 0.05$; right panels) at two different inclinations $\mu = \cos i$. Despite significant model differences, these results compare reasonably well with the corresponding figures from \citet{Poutanen96} and \citet{Schnittman13}.
}
\label{fig:Poutanen}
\end{figure}

Figure \ref{fig:Poutanen} shows the intensity (flux) and ratio $Q/I$ for both cases, at two viewing angles $\mu = \cos i = 0.15$ and 0.55, emitting and tracking 50 (200) packets in the cold (hot) corona model at every cell and timestep cycle, amounting to a total of about a billion packets. Overall, our results agree reasonably well with \citet{Poutanen96} and \citet{Schnittman13}, despite differences in treatments and physics attributes. For example, both earlier publications included photoelectric absorption in their models, and both utilized theoretical approximations for Compton recoil effects. By contrast, our models do not include absorption opacities, but they account for recoil and reflection effects by direct and self-consistent modeling. Despite these differences we observe comparable degrees of polarization, peaking at roughly 15\% and 5\% at $0.1 m_e c^2$ for the cold and hot cases, respectively.

\bibliographystyle{aasjournalv7}



\end{document}